\documentclass[encapsulated]{article} 
\usepackage[encapsulated]{CJK}
\usepackage[hidelinks]{hyperref}  
\usepackage{graphicx}   
\usepackage{amsmath}
\usepackage{hyperref}   
\usepackage{xcolor} 
\usepackage{bm}
\usepackage{upgreek} 
\usepackage{cite}
\usepackage{indentfirst} 
\usepackage{url} 
\usepackage[utf8]{inputenc}
\usepackage{hyperref} 
\usepackage{authblk}

\title{Investigation of magnon behavior in YIG film under microwave excitation using Brillouin light scattering}

\author[1]{Guofu Xu} 
\author[1]{Kang An} 
\author[1]{Wenjun Ma} 
\author[1]{Xiling Li} 
\author[1,2]{C. K. Ong}
\author[1]{Chi Zhang\thanks{Corresponding author. E-mail:~zc@lzu.edu.cn}} 
\author[1]{Guozhi Chai\thanks{Corresponding author. E-mail:~chaigzh@lzu.edu.cn}} 

\affil[1]{Key Laboratory for Magnetism and Magnetic Materials of the Ministry of Education, Lanzhou University, Lanzhou, 730000, China.}
\affil[2]{Department of Physics, Xiamen University Malaysia, Jalan Sunsuria, Bandar Sunsuria, 43900 Sepang, Selangor, Malaysia.}

\date{}  
\begin{document}
\maketitle

\begin{abstract}
We utilize conventional wave-vector-resolved Brillouin light scattering technology to investigate the spin wave response in YIG thin films under high-power microwave excitation. By varying the microwave frequency, external bias magnetic field, and in-plane wave vector, in addition to observing the dipole-exchange spin waves excited by parallel parametric pumping, we further observe broadband spin wave excitation within the dipole-exchange spin wave spectrum. This broadband excitation results from the combined effects of parallel and perpendicular parametric pumping, induced by irregularities in the excitation geometry, as well as magnon-magnon scattering arising from the absence of certain spin wave modes. Our findings offer new insights into the mechanisms of energy dissipation and relaxation processes caused by spin wave excitation in magnetic devices operating at high power.
\end{abstract}



\section{Introduction}

Magnons\cite{magnon1,magnon2,magnon3,magnon4}, or spin waves, feature low-power consumption, are easily tunable, and can be used to develop a new generation of spin-wave logic computing devices\cite{logic1,logic2}. Meanwhile, due to the ease hybridization of magnons, photons, phonons or quantum bits and the presence of spin wave chirality due to spin-orbit coupling, magnon also becomes a potential carrier for the transmission, storage and processing of quantum information\cite{re1,re2}. Consequently, spin wave excitation methods have also been widely studied. Among the various excitation techniques, microwave-excited spin waves are among the most commonly used. Under low-power excitation, spin-wave excitation remains linear. In this regime, the spin waves share the same frequency and wave vector as the driving microwave radiation. However, under high-power microwave excitation, spin waves exhibit not only this linear response mechanism but also nonlinear responses due to their inherent nonlinear properties. This nonlinear process is known as parametric pumping\cite{pp1,pp2,pp3}.

Parametric pumping plays a crucial role in magnetic excitation experiments and applications\cite{ppapplid1}, as it can amplify and process spin wave signals. This technique has been employed to investigate various intriguing phenomena, including spin-wave parametric instability\cite{instab1,instab2}, high-density magnon gases and condensates\cite{gas1,gas2,gas3}, parametric amplification of spin-wave\cite{amp1,amp2,amp3} , and nonlinear magnon logic devices\cite{nd1,nd2}. Among these, one of the most significant achievements is the realization of magnon Bose-Einstein condensation (BEC)\cite{gas3}. Compared to the direct excitation of spin waves, parametric pumping offers two fundamental advantages. First, it enables the excitation of short-wavelength spin waves, which are challenging to achieve with direct excitation due to the antenna width limitation. Second, parametric pumping is a threshold effect—once the threshold is exceeded, the amplification of parametrically excited spin waves grows exponentially. As a result, parametric excitation achieves higher spin-wave excitation efficiency than direct linear excitation.

Here, we use conventional wave-vector-resolved Brillouin light scattering spectroscopy\cite{bls1,bls2,bls3} to directly observe excited spin waves with wave vectors up to 16 rad/$\mu$m under varying pump microwave frequencies and external bias magnetic fields. In addition to detecting the fundamental parallel parametric pumped spin wave signals, we also observe a broad frequency range of spin wave signals within the dipole-exchange spin wave spectrum, along with magnon scattering behavior under high-power excitation.

\section{Experimental details}

In this work, the sample used is a 3.9 $\mu$m-thick YIG single-crystal film, grown on a 0.5 mm-thick GGG substrate via liquid-phase epitaxy, as shown in Fig.1(a). The sample is placed upside down on a 1.143 mm-wide and 3 $\mu m $-thick copper microstrip. Microwaves generated by the microwave source are amplified and transmitted into the microstrip, exciting spin waves in the YIG. A detection laser is focused on the middle of the microstrip to measure the spin-wave signals generated in the YIG. Conventional Brillouin light scattering (BLS) serves as the detection method for spin-wave signals. Leveraging BLS’s inherent wave vector selectivity, spin-wave dynamics are systematically analyzed under varying wave vectors, microwave excitation frequencies, and external bias magnetic fields. The Brillouin light scattering setup and its operating principle are illustrated in Fig.1(b) and 1(c).

The incident laser undergoes inelastic scattering with magnons in the sample, preserving both energy and momentum during the process. The backscattered light is collected and directed into a three-channel Fabry-Pérot interferometer for energy or frequency analysis. The momentum (or wave vector) is determined by the angle $\theta$ between the incident laser direction and the film’s normal, given by $k_{//} = {4\pi \sin\theta}/{\lambda}$, where $\lambda$ = 532 nm is the laser wavelength. In our experiment, the maximum wave vector reaches 16 rad/$\mu$m. To enhance wave vector accuracy, we use an NA = 0.16 lens to focus the laser on the sample and collect the backscattered light at a smaller collection angle\cite{2e1,2e2}. Additionally, the laser power was always 40 mW throughout the experiments.

\begin{center}
\includegraphics[width=0.9\textwidth]{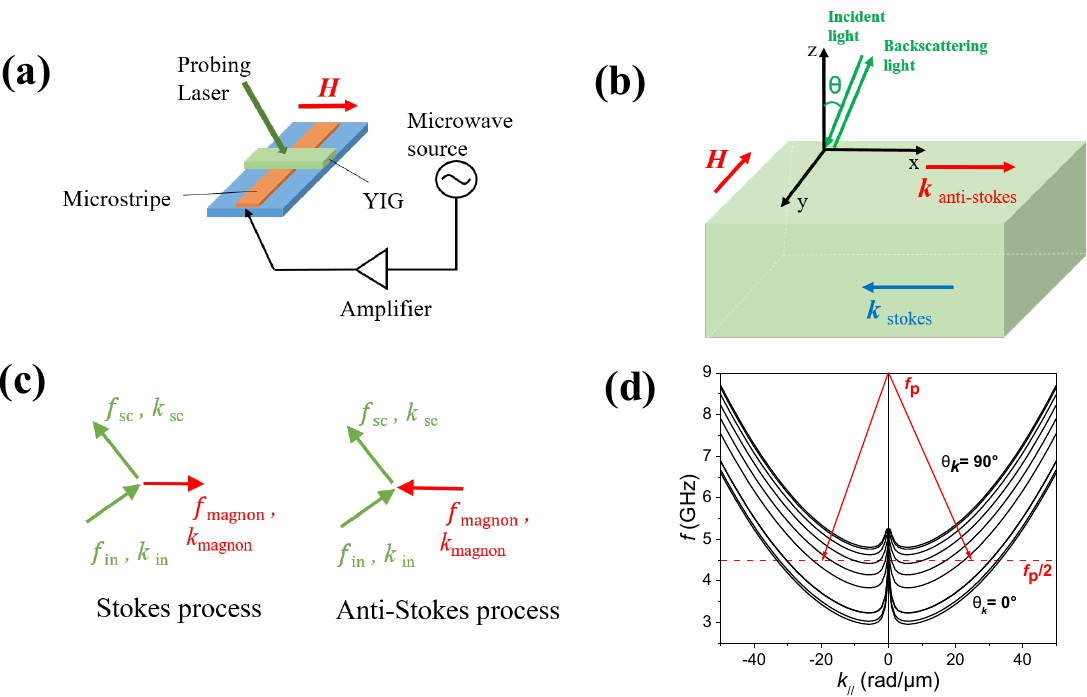}\\[5pt]  
\parbox[c]{\textwidth}{\footnotesize{\bf Fig.~1.}   (a) Schematic illustration of the experimental setup.(b) Schematic diagram of conventional wave-vector-resolved Brillouin light scattering and the inelastic scattering process between magnons and photons(c). (d) Sketch of the lowest-order dipole-exchange spin-wave spectrum(calculated by Eq.(5)) of a 3.9-$\mu$m-thick YIG film under an external field of 1000 Oe and the mechanism of the parametric excitation. Different modes of spin waves represent different $\theta_k$, where $\theta_k$ is the angle between the in-plane wave vector $\mathbf{k}_{\parallel}$ and the external bias magnetic field.}
\end{center}

The external bias magnetic field is aligned parallel to the microwave magnetic field generated by the microstrip line and remains perpendicular to the detected spin wave’s wave vector. This configuration corresponds to the classic Damon-Eshbach (DE) mode \cite{de,de1}. To excite spin waves in the YIG, amplified microwaves at a frequency $f_p$, exceeding the nonlinear threshold power of the sample, are injected into the microstrip. As a result of Suhl instability, two magnetic oscillators with frequency $f_p/2$ appear in the YIG at wave vectors $\pm k$, as depicted in Fig. 1(d), illustrating the parallel parametric pumping process.

\section{Results and Discussion}

\subsection{Parametric pumping}

The nonlinear behavior of ferromagnets was first observed in microwave magnetic resonance experiments in the early 1950s \cite{p6,p7}. Subsequently, Harry Suhl \cite{suhl1} developed a theory based on the parametric generation of spin waves to explain these intriguing observations. Later, Morgenthaler and Schlömann \cite{10,11} theoretically predicted that spin waves could also be excited by a microwave field applied parallel to the static magnetic field. This phenomenon, known as parallel pumping, differs from the perpendicular pumping process observed in the ferromagnetic resonance (FMR) configuration. However, since these are  nonlinear processes, a significant number of magnons are generated only when their production rate exceeds the relaxation rate. Thus, this process is characterized by a threshold pumping field, above which the number of magnons grows exponentially, leading to spin-wave instabilities.

The threshold magnetic fields for parallel pumping(Eq.(1)) and perpendicular pumping(Eq.(2)) are given by the following analytical expressions, respectively:

\begin{equation}
\left(h_{\mathrm{crit}}\right)_{parallel} = \min\left\{\frac{\omega_\mathrm{p} \Delta H_{\mathrm{k}}}{\omega_\mathrm{M} \sin^2 \theta_{\mathrm{k}}}\right\},
\label{eq1} 
\end{equation}

\begin{equation}
\left(h_{\mathrm{crit}}\right)_{perpendicular} = \min\left\{\frac{\omega_{\mathrm{p}} \Delta H_{\mathrm{k}}}{\omega_{\mathrm{M}} \sin 2\theta_k} \right\},
\label{eq2}
\end{equation}

\noindent 
where $\Delta H_k$ is the spin-wave linewidth, ${\omega_\mathrm{M}} = {\gamma 4\pi M_s}$, and $\theta_k$ is the angle between the propagation direction of the parametrically amplified spin waves (wave vector $k$) and the external bias magnetic field. Here, $\gamma$ represents the gyromagnetic ratio, and $M_s$ represents the saturation magnetization.

 When a microwave magnetic field is applied parallel to the static field, for spin waves propagating at an angle to the magnetic field, dipole interactions induce ellipticity in the magnetization precession. As a result, in a spin wave with frequency $\omega_k$, the z-component of magnetization does not remain constant but oscillates over time at a frequency of $2\omega_k$. Consequently, spin waves can be excited at half the driving frequency\cite{magnon4,pp1}.

When the microwave field is perpendicular to the static field, two magnon instability processes—the first-order and second-order Suhl processes—occur at high power levels. Both processes involve magnon interactions: The first-order Suhl process involves three-magnon interactions, where a microwave field at frequency $\omega_p$ generates two magnons with frequency $\omega_p/2$ and wave vectors $\pm k$. The second-order Suhl process involves four-magnon interactions, where a microwave field at frequency $\omega_p$ produces two magnons with frequency $\omega_p$ and wave vectors $\pm k$\cite{magnon4,suhl1}.

\subsection{Spin wave dispersion spectrum in YIG film}

For the dipole exchange spin wave in an infinite medium, its form is derived by Herring and Kittel\cite{kittle}:

\begin{equation}
\omega_{k}^{2} = \left( \omega_{H} + \frac{2A}{M_{s}} \, k^{2} \right) \left( \omega_{H} + \frac{2A}{M_{s}} \, k^{2} + \omega_{M} \sin^{2} \theta_{k} \right),
\end{equation}

\noindent 
where $\omega_H=\gamma\mu_0H_{ext}$, ${\omega_\mathrm{M}}={\gamma4\pi\mathrm{M}_s}$, $\theta_k$  is the angle between the direction of the wavevector $k$ and the direction of the saturation magnetization $M_s$, and $A$ is the exchange stiffness.
  
In our YIG film with finite thickness $d$, due to the boundary conditions of the film, the perpendicular component  of the spin wave vector $k$ becomes quantized:

\begin{equation}
k^2= k_{\parallel}^2 + k_{\perp}^2=k_{\parallel}^2 + ({\frac{n\pi}{d}})^2 
\quad   n = 0, 1, 2, \dots,
\end{equation}

\noindent 
where $k_{\parallel}$ is the magnitude of in-plane wave-vector $k$, $k_{\perp}$ is the magnitude of  perpendicular wave-vector $k$, $d$ is the thickness of film, in our experiment, the thickness of the YIG film is 3.9 $\mu$m, and the impact is much smaller than the resolution of our BLS test, which is generally 50 MHz, therefore, in this article, only the lowest order dipole exchange spin wave, that is, the mode of n=0, is discussed.

For this case, the dipole exchange spin wave spectrum is derived by Kalinikos and Slavin\cite{ks}:

\begin{equation}
\boldsymbol{\omega}_k^2 = \left( \boldsymbol{\omega}_H + \frac{2A}{M_s} \, k^2 + \boldsymbol{\omega}_M \left(1 - F_0 \right) \right) \left( \boldsymbol{\omega}_H + \frac{2A}{M_s} \, k^2 + \boldsymbol{\omega}_M F_0 \sin^2 \theta_k \right),
\end{equation}

\noindent where

\begin{equation}
F_0 = 1-\frac{1 - \exp(-k_{\parallel}d)}{k_{\parallel}d}.
\end{equation}

 Fig.1(d) shows the lowest-order mode dipole-exchange spin waves calculated by Eq.(5) of a 3.9 $\mu$m-thick YIG film under an external magnetic field of 1000 Oe, where different modes represent different $\theta_k$.

\subsection{Observation of parametric pumping}

In our experiment, we fixed the microwave excitation power, where the microwave source generated a signal with a power of -3 dBm. After amplification by a 30 dB gain amplifier, the signal was fed into the microstrip line to excite the YIG sample. The microwave frequency $f_p$ was scanned from 2 GHz to 10 GHz with a step size of 50 MHz. At each frequency step, the Brillouin backscattered light was collected for 50 scans. As shown in Fig.2(a), the original Brillouin spectrum was obtained under an external bias magnetic field of 328 Oe and a detection wave vector of 4.10 rad/$\mu$m. When the microwave excitation frequency $f_p$=4.8 GHz, that is, $f_p/2$ exactly satisfies the spin wave dispersion relation (Eq.(5)), the extremely strong parametrically excited spin wave signal can be observed compared with other microwave excitation frequencies (such as $f_p$=4 GHz or $f_p$=5 GHz) (note the logarithmic scale). This is because under the parametric process, the number of parametrically excited magnons increases exponentially compared to when Eq.(5) is not satisfied. This process is highlighted by the red arrow in Fig. 2(c).

By varying the microwave frequency, we obtained the relationship between the spin wave frequency and the microwave frequency at a fixed wave vector $k_{\parallel}$=4.10 rad/$\mu$m and external bias magnetic field, as shown in Fig. 2(b). The parametric excitation signal is only observed at specific frequencies. Unlike Reference\cite{unlike} , where both parametric and linear excitation signals were observed simultaneously, our experiment—using conventional BLS with a small collection angle and high wave vector resolution—only detects spin waves when the frequency strictly satisfies the spin wave dispersion relation.

We categorize the observed signals into two regions: the first corresponds to direct excitation at $\theta_k=90^\circ$, represented by the strongest red area in Fig. 2(b), while the second spans from 0 to 90 degrees in the middle of the spectrum, visible over a broader frequency range and marked by the red arrow in Fig. 2(c). The parametric processes governing these two regions differ. The spin wave signal excited in the first process is significantly stronger than in the second, which could be attributed to two factors. First, in our experimental configuration of parallel parametric pumping, where the applied microwave field is parallel to the bias magnetic field, the parametric excitation efficiency above the threshold is strongly dependent on $\theta_k$. As indicated by the Eq.(1), the excitation efficiency is maximized when $\theta_k=90^\circ$. Second, our BLS detection setup ensures that the test wave vector $k_{\parallel}$ remains perpendicular to the applied magnetic field, achieving the highest detection efficiency when $\theta_k=90^\circ$.

The second process may not only be generated by parallel parametric pumping, but also by perpendicular parametric pumping. Due to the limited width of the microstrip line, there will be a microwave magnetic field perpendicular to the external bias magnetic field at the edge of the microstrip line, which may lead to perpendicular parametric pumping\cite{filed1,filed2}. The spin wave excitation in this broad area may come from the combined effect of the two parametric excitations.

\begin{center}
\includegraphics[width=0.8\textwidth]{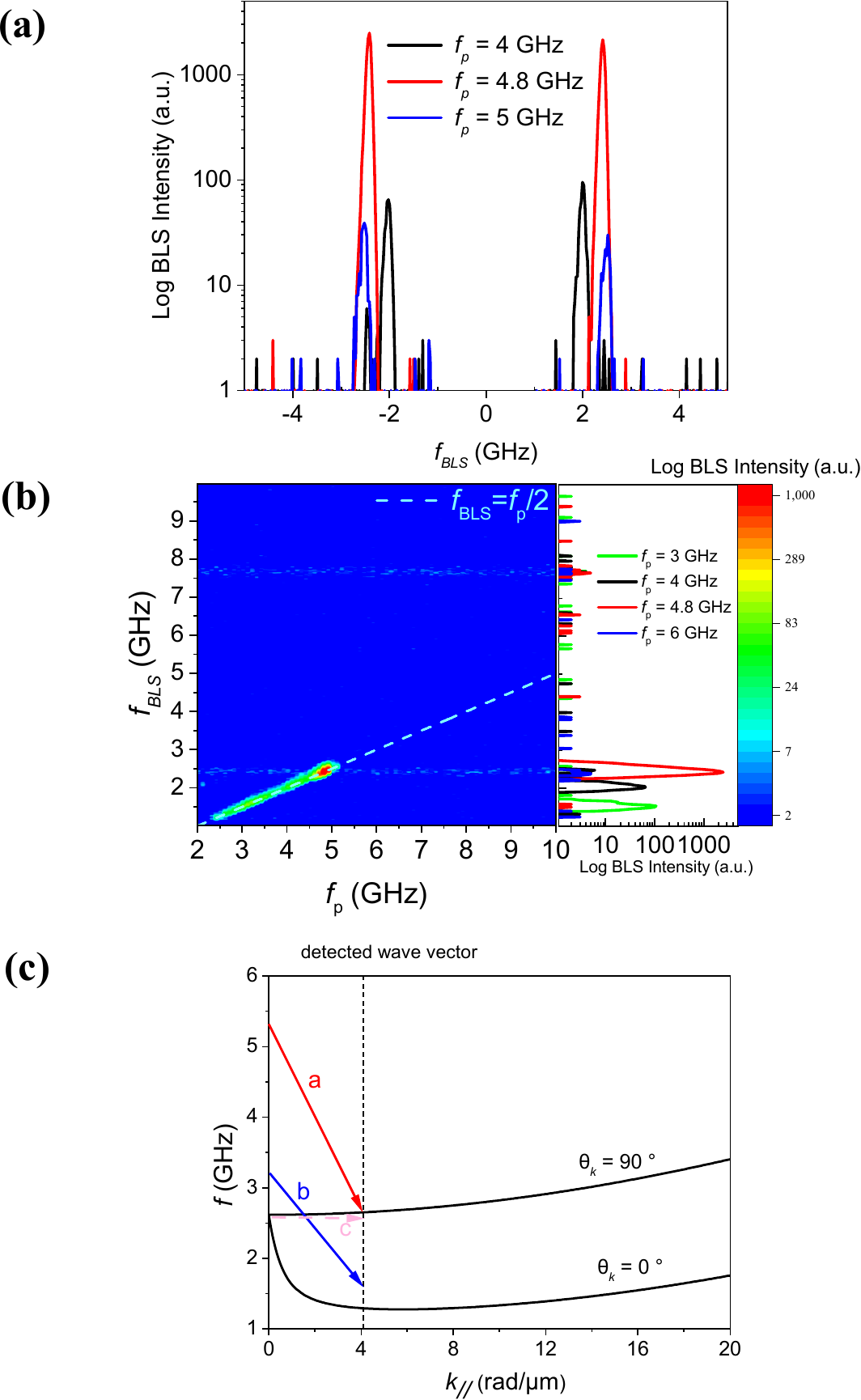}\\[5pt]  
\parbox[c]{\textwidth}{\footnotesize{\bf Fig.~2.} (a) BLS original spectra under three different frequencies of microwave excitation at a bias magnetic field of 328 Oe and a detection wave vector of 4.10 rad/$\mu$m. (b) The relationship between the parametric excited spin wave frequency and the microwave excitation frequency, the color scale represents the BLS intensity. (c) Three different spin wave generation processes under high-power microwave excitation.}
\end{center}

\subsection{Magnon interaction of excited spin wave}

We now study how the signal of nonlinear parametrically excited spin waves in the YIG sample changes as the external bias magnetic field varies. We set the wave vector of the detected spin wave to 4.10 rad/$\mu$m, adjust the external bias magnetic field, and obtain a series of graphs similar to Fig. 2(b), showing the relationship between spin wave frequency and microwave frequency. We then convert these into an animated GIF, as shown in Supplementary Material 1.

To better reflect the relationship between the parametrically excited spin wave and the external magnetic field, we integrate each BLS original spectrum, and the result represents the number of magnons corresponding to the measured wave vector under a given magnetic field and microwave excitation frequency. In this way, we obtain a mapping diagram of microwave frequency versus external magnetic field, as shown in Fig.3, where the color scale represents the number of magnons.

\begin{center}
\includegraphics[width=0.9\textwidth]{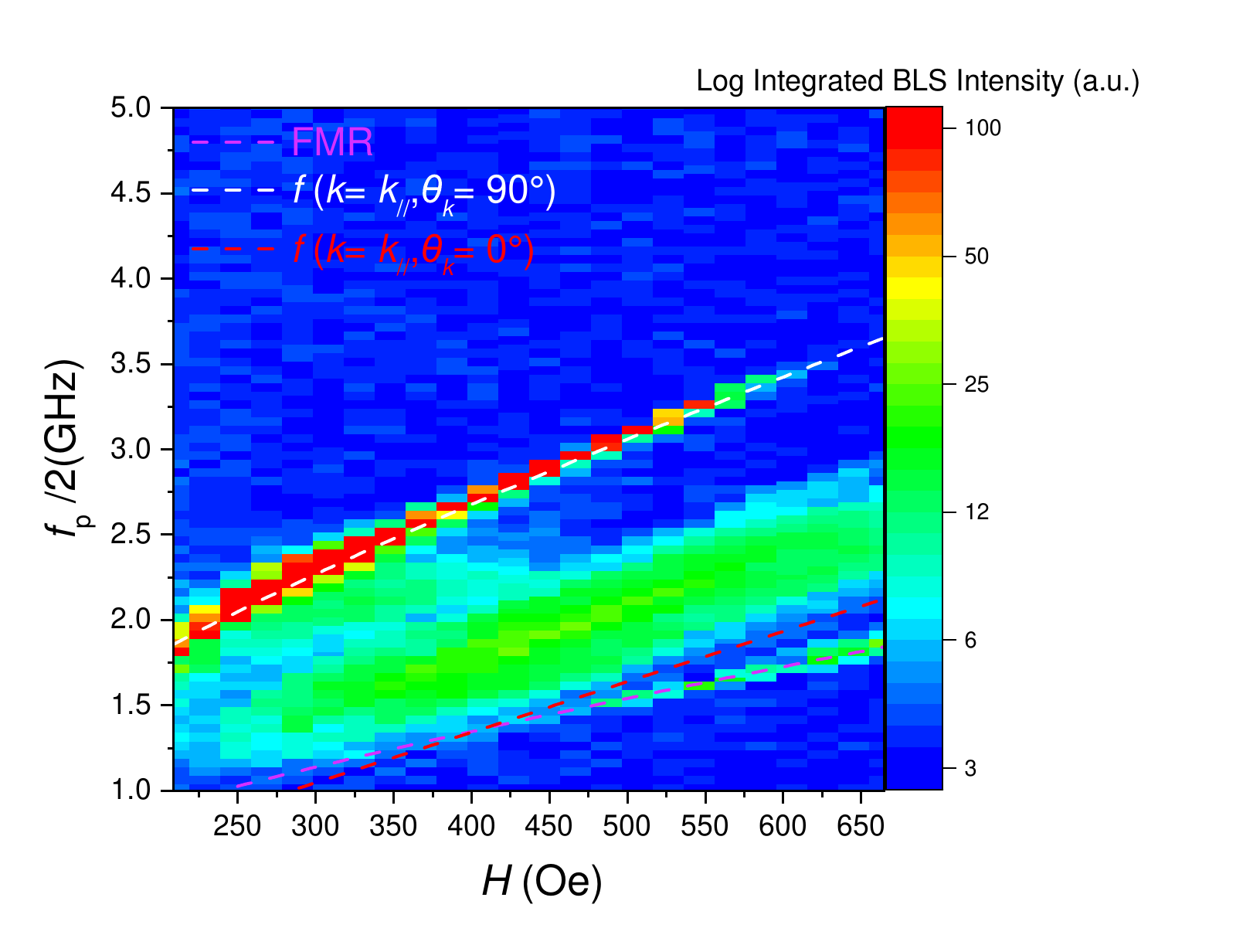}\\[5pt]  
\parbox[c]{\textwidth}{\footnotesize{\bf Fig.~3.}   Mapping of microwave excitation frequency and applied bias magnetic field at in-plane wave vector $k_{\parallel}$=4.10 rad/$\mu$m. The color scale represents the integrated BLS intensity, which represents the number of magnons.}
\end{center}

We observe three distinct excitation processes. Process $a$ corresponds to the first excitation process mentioned above, where the spin wave mode with $\theta_k=90^\circ$ is excited. This process is the most important parametric process, parallel parametric pumping process. As the external magnetic field increases, this process gradually weakens and then suddenly disappears. This is attributed to the fact that as the bias magnetic field increases,  both the microwave frequency $f_p$ and the spin wave linewidth $\Delta H_{\mathrm{k}}$ increase, as shown in Eq.(1). Beyond a certain magnetic field, the microwave magnetic field can no longer reach the threshold required for parallel parametric excitation, causing the parametric spin wave signal to abruptly vanish, which is a typical characteristic of nonlinear effects.

Process b corresponds to the second excitation process mentioned above. It is a parametric excitation process occurring within the broadband spin wave spectrum, which includes multiple modes within $0^\circ \leq \theta_k < 90^\circ$. In parallel parametric pumping, the coupling efficiency between RF microwaves and spin waves follows a $\sin^2\theta_k$  relationship, which peaks at $90^\circ$, while in perpendicular parametric pumping, it follows a $\sin2\theta_k$ relationship. This broadband excitation represents a hybrid excitation process.
Process c is a distinct mode not observed in the previous cases, and it only emerges as the magnetic field increases. Unlike processes a and b, this is a same-frequency excitation process, where the excited spin wave frequency matches the ferromagnetic resonance (FMR) frequency under the given magnetic field, rather than following the dipole-exchange spin wave modes.

\begin{center}
\includegraphics[width=0.9\textwidth]{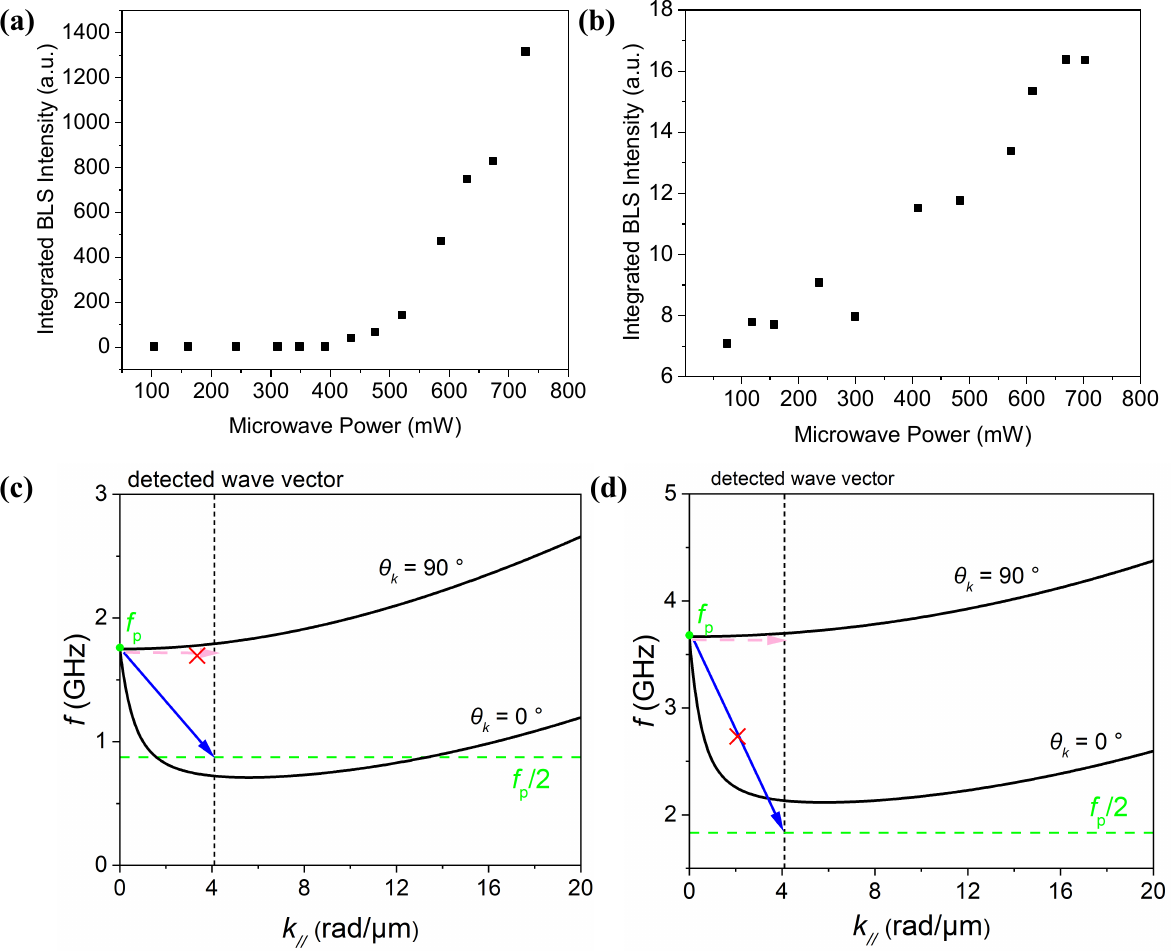}\\[5pt]  
\parbox[c]{\textwidth}{\footnotesize{\bf Fig.~4.}   BLS intensity relationship under different microwave powers, corresponding to Process $a$ (a) and Process $c$ (b). Dipolar-exchange spin wave dispersion and spin wave generation process of Process $c$ under different external bias magnetic fields, $H$=200 Oe (c), $H$=700 Oe (d).}
\end{center}

To further understand this process, we conducted variable power tests on processes a and c, as shown in Figs.4(a) and 4(b), respectively. In Process $a$, the spin wave intensity and microwave power exhibit a clear nonlinear relationship, further confirming its parametric pumping nature. In contrast, Process $c$ shows an approximately linear relationship, similar to spin wave excitation under low-power conditions.

Additionally, we found that Process $c$ does not occur unless the FMR frequency exceeds the parametric excitation spin-wave frequency. This can be understood through the dispersion relation illustrated in Figs.4(c) and 4(d). At low magnetic fields, both parametric process and FMR can occur, and parametric pumping is more likely to occur at this time. However, at high magnetic fields, microwave excitation with the frequency $f_p$ cannot cause parametric process because there is no spin wave mode at frequency $f_p/2$. Instead, over-accumulation magnons gather at $k = 0$ and scatter into other wave vectors, as indicated by the pink dashed arrow in Fig.2(c).

Finally, we attribute Process $c$ to the perpendicular microwave magnetic field generated at the edges of the microstrip line, which excites $k = 0$ magnons(FMR). These magnons, unable to relax efficiently, undergo either double-magnon scattering or four-magnon scattering. As a result, we observe spin wave signals in the BLS spectrum that match the microwave frequency.

\subsection{Parametric excitation under different wave vectors}

We also investigated spin wave signals under different wave vectors during parametric excitation. Based on the above research , we varied the wave vector and conducted experiments under the same conditions. The relationship between the original BLS spectrum and the magnetic field is shown in Supplementary Materials 1-4., while the integrated BLS spectrum is presented in Figs.5(a)–(d), corresponding to wave vectors of 4.10, 8.08, 11.81, and 15.18 rad/$\mu$m, respectively.

We found that as the wave vector increases, the parametric process follows the dipole-exchange spin wave dispersion relation described in Eq.(5). The three excitation processes (a, b and c) are also observed at different wave vectors. However, the intensity of parametric excitation gradually decreases with increasing wave vector. This is expected, as a larger wave vector leads to an increase in spin wave linewidth (or magnon dissipation rate) and a decrease in magnon scattering probability, thereby reducing the efficiency of microwave-driven spin wave excitation , eventually weakening the parametrically excited spin wave signal.

\begin{center}
\includegraphics[width=0.9\textwidth]{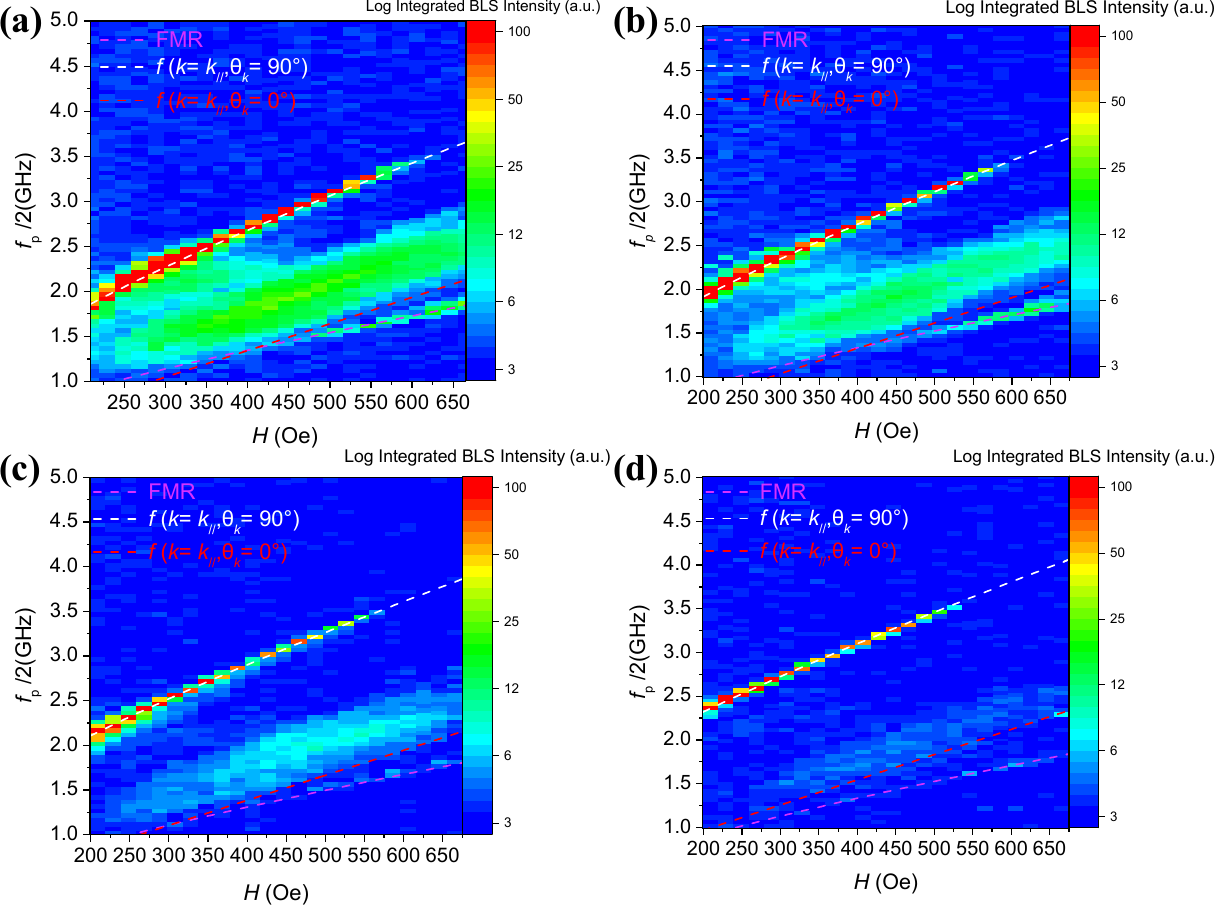}\\[5pt]  
\parbox[c]{\textwidth}{\footnotesize{\bf Fig.~5.}   The mapping diagram of microwave excitation frequency and external bias magnetic field under different in-plane wave vectors, corresponding to $k_{\parallel}$= 4.10 rad/$\mu$m (a), $k_{\parallel}$= 8.08 rad/$\mu$m (b), $k_{\parallel}$= 11.81 rad/$\mu$m (c), $k_{\parallel}$= 15.18 rad/$\mu$m (d).}
\end{center}

\section{Conclusion}

In summary, we systematically studied the excitation of spin waves in YIG films under high-power microwave excitation. Our observations of spin waves with wave vectors up to 16 rad/$\mu$m under various pump microwave frequencies and external bias magnetic fields reveal that, in addition to the fundamental parallel parametric pumping process, we also observe broadband parametric excitation within the dipole-exchange spin wave spectrum, as well as magnon scattering effects under high-power excitation. These findings offer new insights into energy dissipation and relaxation mechanisms in magnetic devices operating under high power.

\

\addcontentsline{toc}{chapter}{Acknowledgment}
\section*{Acknowledgment}
This work is supported by the National Natural Science Foundation of China (NSFC) (No. 52471200, 12174165 and 52201219).

\addcontentsline{toc}{section}{References}
\footnotesize
\bibliographystyle{iopart-num} 
\bibliography{references} 

\providecommand{\newblock}{}
\begin{thebibliography}{10}
\expandafter\ifx\csname url\endcsname\relax
  \def\url#1{{\tt #1}}\fi
\expandafter\ifx\csname urlprefix\endcsname\relax\def\urlprefix{URL }\fi
\providecommand{\eprint}[2][]{\url{#2}}

\bibitem{magnon1}
Neusser S and Grundler D 2009 {\em Advanced Materials\/} {\bf 21} 2927--2932

\bibitem{magnon2}
Kruglyak V~V, Demokritov S~O and Grundler D 2010 {\em Journal of Physics D: Applied Physics\/} {\bf 43} 264001

\bibitem{magnon3}
Chumak A~V, Vasyuchka V~I, Serga A~A and Hillebrands B 2015 {\em Nature Physics\/} {\bf 11} 453--461

\bibitem{magnon4}
Rezende S~M 2020 {\em Fundamentals of magnonics\/} vol 969 (Springer)

\bibitem{logic1}
Chumak A~V, Kabos P, Wu M, Abert C, Adelmann C, Adeyeye A~O, {\AA}kerman J, Aliev F~G, Anane A, Awad A {\em et~al.\/} 2022 {\em IEEE Transactions on Magnetics\/} {\bf 58} 1--72

\bibitem{logic2}
Schneider T, Serga A~A, Leven B, Hillebrands B, Stamps R~L and Kostylev M~P 2008 {\em Applied Physics Letters\/} {\bf 92} 022505

\bibitem{re1}
Yuan H, Cao Y, Kamra A, Duine R~A and Yan P 2022 {\em Physics Reports\/} {\bf 965} 1--74

\bibitem{re2}
Yu T, Luo Z and Bauer G~E 2023 {\em Physics Reports\/} {\bf 1009} 1--115

\bibitem{pp1}
Bracher T, Pirro P and Hillebrands B 2017 {\em Physics Reports\/} {\bf 699} 1--34

\bibitem{pp2}
Schl{\"o}mann E, Green J and Milano u 1960 {\em Journal of Applied Physics\/} {\bf 31} S386--S395

\bibitem{pp3}
Rezende S~M and de~Aguiar F~M 1990 {\em Proceedings of the IEEE\/} {\bf 78} 893--908

\bibitem{ppapplid1}
Pirro P, Vasyuchka V~I, Serga A~A and Hillebrands B 2021 {\em Nature Reviews Materials\/} {\bf 6} 1114--1135

\bibitem{instab1}
De~Aguiar F and Rezende S 1986 {\em Physical review letters\/} {\bf 56} 1070

\bibitem{instab2}
Azevedo A and Rezende S~M 1991 {\em Physical review letters\/} {\bf 66} 1342

\bibitem{gas1}
Bozhko D~A, Clausen P, Melkov G~A, L’vov V~S, Pomyalov A, Vasyuchka V~I, Chumak A~V, Hillebrands B and Serga A~A 2017 {\em Physical review letters\/} {\bf 118} 237201

\bibitem{gas2}
Bozhko D~A, Serga A~A, Clausen P, Vasyuchka V~I, Heussner F, Melkov G~A, Pomyalov A, L’vov V~S and Hillebrands B 2016 {\em Nature Physics\/} {\bf 12} 1057--1062

\bibitem{gas3}
Demokritov S~O, Demidov V~E, Dzyapko O, Melkov G~A, Serga A~A, Hillebrands B and Slavin A~N 2006 {\em Nature\/} {\bf 443} 430--433

\bibitem{amp1}
Breitbach D, Schneider M, Heinz B, Kohl F, Maskill J, Scheuer L, Serha R~O, Br{\"a}cher T, L{\"a}gel B, Dubs C {\em et~al.\/} 2023 {\em Physical Review Letters\/} {\bf 131} 156701

\bibitem{amp2}
Serga A, Hillebrands B, Demokritov S, Slavin A, Wierzbicki P, Vasyuchka V, Dzyapko O and Chumak A 2005 {\em Physical review letters\/} {\bf 94} 167202

\bibitem{amp3}
Verba R, Carpentieri M, Finocchio G, Tiberkevich V and Slavin A 2018 {\em Applied Physics Letters\/} {\bf 112} 042402

\bibitem{nd1}
Ge X, Verba R, Pirro P, Chumak A~V and Wang Q 2024 {\em Applied Physics Letters\/} {\bf 124} 122413

\bibitem{nd2}
Ustinov A~B, L{\"a}hderanta E, Inoue M and Kalinikos B~A 2019 {\em IEEE Magnetics Letters\/} {\bf 10} 1--4

\bibitem{bls1}
Wilber W, Wettling W, Kabos P, Patton C and Jantz W 1984 {\em Journal of Applied Physics\/} {\bf 55} 2533--2535

\bibitem{bls2}
Sandweg C, Jungfleisch M, Vasyuchka V, Serga A, Clausen P, Schultheiss H, Hillebrands B, Kreisel A and Kopietz P 2010 {\em Review of Scientific Instruments\/} {\bf 81} 073902

\bibitem{bls3}
Serga A, Sandweg C, Vasyuchka V, Jungfleisch M, Hillebrands B, Kreisel A, Kopietz P and Kostylev M 2012 {\em Physical Review B—Condensed Matter and Materials Physics\/} {\bf 86} 134403

\bibitem{2e1}
Demokritov S~O, Hillebrands B and Slavin A~N 2001 {\em Physics Reports\/} {\bf 348} 441--489

\bibitem{2e2}
Sandercock J and Wettling W 1973 {\em Solid State Communications\/} {\bf 13} 1729--1732

\bibitem{de}
Eshbach J and Damon R 1960 {\em Physical Review\/} {\bf 118} 1208

\bibitem{de1}
Damon R~W and Eshbach J 1961 {\em Journal of Physics and Chemistry of Solids\/} {\bf 19} 308--320

\bibitem{p6}
Bloembergen N and Damon R 1952 {\em Physical Review\/} {\bf 85} 699

\bibitem{p7}
Damon R~W 1953 {\em Reviews of Modern Physics\/} {\bf 25} 239

\bibitem{suhl1}
Suhl H 1957 {\em Journal of Physics and Chemistry of Solids\/} {\bf 1} 209--227

\bibitem{10}
Morgenthaler F~R 1960 {\em Journal of Applied Physics\/} {\bf 31} S95--S97

\bibitem{11}
Schl{\"o}mann E, Green J and Milano u 1960 {\em Journal of Applied Physics\/} {\bf 31} S386--S395

\bibitem{kittle}
Herring C and Kittel C 1951 {\em Physical Review\/} {\bf 81} 869

\bibitem{ks}
Kalinikos B and Slavin A 1986 {\em Journal of Physics C: Solid State Physics\/} {\bf 19} 7013

\bibitem{unlike}
Schultheiss H, Janssens X, van Kampen M, Ciubotaru F, Hermsdoerfer S, Obry B, Laraoui A, Serga A, Lagae L, Slavin A {\em et~al.\/} 2009 {\em Physical review letters\/} {\bf 103} 157202

\bibitem{filed1}
Neumann T, Serga A, Vasyuchka V and Hillebrands B 2009 {\em Applied Physics Letters\/} {\bf 94} 192502

\bibitem{filed2}
Wang P 2023 {\em Chinese Physics B\/} {\bf 32} 037601

\end{thebibliography}

\end{document}